\documentclass[preprint,12pt]{elsarticle}
\usepackage{graphicx}
\usepackage{amssymb}
\usepackage{amsmath}
\journal{Physics Letters  B}
\begin{document}
\begin{frontmatter}
\title{Effects of massive photons from the dark sector on the muon content
in extensive air showers}
\author{Jan Ebr\corref{cor1}}
\ead{ebr@fzu.cz}
\cortext[cor1]{Corresponding author}
\author{Petr Ne\v{c}esal\corref{}}
\ead{necesal@fzu.cz}
\address{Institute of Physics of the Academy of Sciences of the Czech Republic, Na Slovance 1999/2, 18221 Prague 8, Czech
Republic}
\begin{abstract}
Inspired by recent astrophysical observations of leptonic excesses measured by satellite experiments, we consider the impact of some general models of the dark sector on the muon production in extensive air showers. We present a compact approximative expression for the bremsstrahlung of a massive photon from an electron and use it within Monte Carlo simulations to estimate the amount of weakly interacting photon-like massive particles that could be produced in an extensive air shower. We find that the resulting muon production is by many orders of magnitude below the average muon count in a shower and thus unobservable.
\end{abstract}
\end{frontmatter}

\section{Motivation}

The relatively recent observations of excess lepton fluxes from space,
as measured by PAMELA \cite{Adriani:2008zr} and ATIC \cite{2008Natur.456..362C}
have motivated large interest in models of dark matter annihilation
that could explain these data, while staying in agreement with other existing
astrophysical evidence. For the ultra-high energy
cosmic ray (UHECR) experiments, neither these low-energy fluxes, nor
their hypothetical parent particles are directly observable. But the
common feature of such models is that they need to add new physics
to increase the production of leptons with respect to hadrons in the
current universe. While this production is mainly targeted at electrons
and positrons, many of such processes also lead to extra production
of muons. 

Some evidence of disagreement in muon production in extensive air showers
initiated by high-energy cosmic rays with the predictions of Monte Carlo simulations
has been given already by the DELPHI \cite{Abdallah:2007fk}, ALEPH \cite{aleph} and L3 \cite{L3} experiments at LEP 
and it has been repeatedly
reported by the Pierre Auger Observatory \cite{PierreAuger:2011aa}.
In both cases, the data indicate that the current interaction models
may significantly underestimate the number of muons produced.

 It is
then only natural to ask, whether some of these extra muons could
be accounted for if some of the above mentioned new physics is incorporated
into the Monte Carlo simulations. Instead of considering every single
model of the dark sector that has ever been proposed (for a very recent review of models with specific references, see Ch. 4 of \cite{DMrev}), we turn to the
work \cite{ArkaniHamed:2008qn}, which is rather general. There it
is argued that, considering not only the above-mentioned excess, but
also general cosmological observations and direct dark matter searches,
it is not unreasonable to expect the TeV-scale dark matter to be accompanied
by a relatively light particle with mass around 250 MeV and some weak coupling
to ordinary matter. This idea is further corroborated in \cite{Cheung:2009qd}
for the special case of such a particle being essentially a massive
photon that couples to ordinary matter via kinetic mixing suppressed
by a small factor of the order of $\epsilon\approx10^{-2}-10^{-3}$. We call
this particle a ``dark photon" for brevity as the factor $\epsilon$ effectively appears in any vertex that includes both a standard model particle and a dark photon, thus making it difficult to detect by electromagnetic interactions.
This scheme is not only backed by a compelling theoretical motivation,
but also relatively simple to implement as a first look into the topic,
yet reasonably general; thus we focus on it in the rest of the paper. 

In the following article \cite{Cheung:2009su} the authors show that
this model leads to the prediction of specific collider signatures
in the form of ``lepton jets" stemming
from the prediction of the TeV-scale particles in hadronic collisions.
These events are too rare to have any effect on extensive air showers,
as there are only hundreds of sufficiently high-energy hadronic interactions
in a single air shower, which itself is a rather rare event -- the
total luminosity in UHECRs is simply too small for even Standard model
electroweak effects to have any impact on observable data, even more
so for exotics. Nevertheless, this model is still interesting because
the dark photon, being coupled to ordinary electric charge, albeit
weakly, can be produced via bremsstrahlung from electrons. The amount
of electromagnetic interactions of photons and leptons in each shower
is by many orders of magnitude larger than that of hadronic interactions
and so it is not immediately obvious what size of cross-section for the dark
photon bremsstrahlung (investigated in the next section) is needed
to produce observable effects.

The attractive feature of a massive photon is that it can decay into
a pair of a charged particle and its antiparticle. Additionally, the
dark photon is ``dark", that is, it has limited interactions with ordinary matter. Thus, dark photons in the relevant
range of masses will almost always decay instead of producing a pair in the electromagnetic field of an atom in the air. In
the case of pair production, almost all produced pairs are electron-positron
as the cross-section falls with the fourth power of the lepton mass, whereas 
the decay of the dark photon proceeds democratically into every kinematically possible
final state, save for threshold effects. Thus, for mass of the dark photon $m_{\gamma}\in(212,280)$
MeV, for every dark photon produced, there is on average one muon
added to the shower. For higher masses, pion final states are possible,
but muon production is still sizeable.

\section{Dark photon bremsstrahlung}

The problem of bremsstrahlung of a massless photon from a lepton interacting
with an atomic target in quantum electrodynamics is a well-known one
and, to the leading order, it is exhaustively described in \cite{Tsai:1973py}.
Interestingly, we did not find an expression for the bremsstrahlung
of a massive photon in any literature, so we had to derive one ourself. Elementary
as it may seem, the calculation is actually quite tedious. Thus,
even though it is technically possible to just add a photon mass into
the equations in \cite{Tsai:1973py} and proceed, this would be a
major task and prone to errors. Instead we note the work \cite{Kim:1973he}
where it is shown that similar results can be derived using the computationally
much simpler Weizs\"{a}cker-Williams approximation, where the 2$\rightarrow$3
problem is reduced to a 2$\rightarrow$2 Compton scattering times
a factor determined by kinematics and the scattering target. Schematically 
\begin{equation}
\frac{d\sigma\left(2\rightarrow3\right)}{d\left(P_{1}\cdot k\right)d\left(P_{i}\cdot k\right)}=\frac{d\sigma\left(2\rightarrow2\right)}{d\left(P_{1}\cdot k\right)}_{t=t_{min}}\frac{\alpha}{\pi}\frac{\chi}{P_{2}\cdot P_{i}},\label{eq:WW}
\end{equation}
where $P_{i}$ is the initial four-momentum of the target, $P_{1}$
and $P_{2}$ are the initial and final four-momenta of the lepton,
$k$ is the four-momentum of the produced (massive or not) photon, $\alpha=e^2/4\pi$ is the fine structure constant
and $\chi$ is a factor that involves the form-factors of the target,
which is independent of the 2$\rightarrow$2 process. The subscript
$t=t_{min}$ denotes that the 2$\rightarrow$2 process is evaluated
using a particular kinematic set up. 

In \cite{Tsai:1986tx}, this approximation is used for the bremsstrahlung
of a massive axion. While the resulting formula is not directly applicable
to the production of a massive photon, most of the work is actually
done. The difference is only in the matrix element for the 2$\rightarrow$2
process, which is a well-known function. The difficult part,
which is the kinematics, is exactly the same for any massive particle, while being vastly different
from the case of a massless photon. From eq. (7) of \cite{Tsai:1986tx} we observe that the kinematics for the  2$\rightarrow$2
process can be worked out so that 
\begin{equation}
\frac{d\sigma\left(2\rightarrow2\right)}{d\left(P_{1}\cdot k\right)}_{t=t_{min}}=\frac{1}{16\pi\left(P_{2}\cdot k\right)^2}\lvert\overline{\mathcal{M}}\rvert^2,
\end{equation}
where $\lvert\overline{\mathcal{M}}\rvert$ is the absolute value of the invariant matrix element for the 2$\rightarrow$2 process averaged over initial state polarisations and summed over final state polarisations. For massive photons,
\begin{equation}
\lvert\overline{\mathcal{M}}\rvert^2=-16\pi^2\alpha^2\frac{2\left(m_{\gamma}^{4}+2m_{\gamma}^{2}(P_{2}\cdot k-P_{1}\cdot k)+2\left(\left(P_{1}\cdot k\right)^{2}+\left(P_{2}\cdot k\right)^{2}\right)\right)}{\left(m_{\gamma}^{2}-2P_{1}\cdot k\right)\left(m_{\gamma}^{2}+2P_{2}\cdot k\right)},
\end{equation}
where we can safely neglect
the electron mass if we are interested in dark photons capable of
decaying into two muons. This approximation was numerically checked against the
full result and the agreement is better than a fraction of a per cent
for $m_{\gamma}=250$ MeV in almost the whole range of $x$, while
the length of the formula is significantly  reduced. Using eq. (\ref{eq:WW})
and kinematics, we deduce that the bremsstrahlung cross-section is
\begin{equation}
\frac{d\sigma}{dxd\Omega}=\frac{\alpha^{3}E_{1}x\left(x^{2}-2x+2\right)\chi}{\pi\left(m_{e}^{2}x\left(1+\left(\frac{E_{1}}{m_{e}}\right)^{2}\theta^{2}\right)+m_{\gamma}^{2}\left(\frac{1}{x}-1\right)\right)^{2}},
\end{equation}
where $x$ is the fraction of $E_{1}$ carried by the produced dark
photon and $\theta$ is its production angle with respect to the incoming electron in the laboratory frame. Here we must keep the
electron mass non-zero not only because of the behaviour for $x\rightarrow1$
but also because the $\gamma$-factor of the electron can be huge. 

To proceed with the angular integration we must specify the $\chi$-factor.
Again, we take it from \cite{Tsai:1986tx}. In the ``complete
screening" limit it can be written as 
\begin{equation}
\begin{split}
\chi=2\left[Z\ln\left(\frac{1194}{Z^{2/3}}\right)+Z^{2}\ln\left(\frac{184}{Z^{1/3}}\right)+\right. \\ 
\left.+\left(Z+Z^{2}\right)\left(\ln\left(1+\left(\frac{E_{1}}{m_{e}}\right)^{2}\theta^{2}\right)-1\right)\right],
\end{split}
\end{equation}
where $Z$ is the atomic number of the target. {}``Complete screening''
refers to an approximation valid when 
\begin{equation}
\begin{split}
184\,\mathrm{e}^{-1/2}\frac{Z^{-1/3}}{m_e}\,t_{min} \ll1 \\
1194\,\mathrm{e}^{-1/2}\frac{Z^{-2/3}}{m_e}\, t_{min}\ll1
\end{split}
\end{equation}
\textendash{} an explicit numerical calculation again
shows, that it is well justified. To make the production of a 250 MeV dark photon even possible, the $\gamma$-factor squared of the
electron has to be in the order of at least $10^{5}$ and thus the
scattering is strongly suppressed for large angles. The suppression
is in fact strong enough that we can extend the integral in $\theta$
to infinity, yielding a much more compact analytical result, namely

\begin{equation}
\begin{split}
\frac{d\sigma}{dx}=\frac{4\alpha^{3}x((x-2)x+2)}{E_{1}}\times \\
\times\left[\frac{Z+Z^{2}-Z\ln\left(\frac{1194}{Z^{2/3}}\right)-Z^{2}\ln\left(\frac{184}{Z^{1/3}}\right)}{m_{\gamma}^{2}(x-1)-m_{e}^{2}x^{2}}+\right.\\
\left.+\frac{(Z+Z^{2})\log\left(\frac{m_{e}^{2}x^{2}}{m_{\gamma}^{2}(1-x)+m_{e}^{2}x^{2}}\right)}{m_{\gamma}^{2}(x-1)}\right],\label{eq:darkBS}
\end{split}
\end{equation}

Here we keep only the electron masses that cannot be neglected by any means. Our ultimate
goal is to compare this expression with the well-known formula for
massless photon bremsstrahlung, that is 

\begin{figure}
\includegraphics{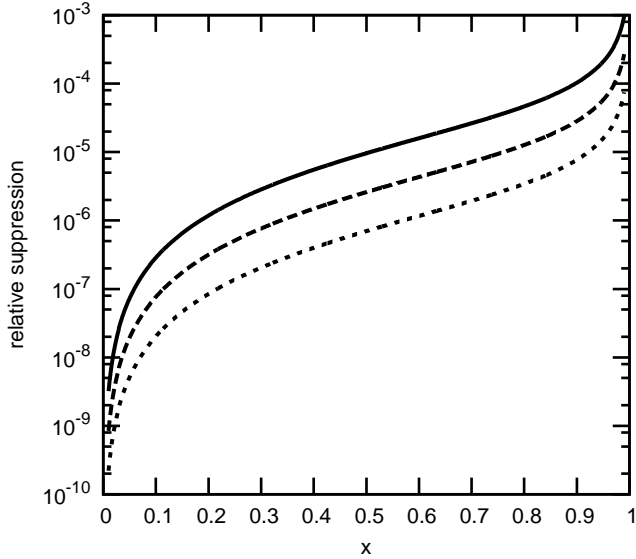}

\caption{\label{fig:250-500-1}The suppression of the bremsstrahlung by the
mass of the photon, relatively to the case of a massless photon for
$Z=7$ and different photon masses: solid 250 MeV, dashed 500 MeV, dotted
1 GeV, in dependence on the transferred energy fraction.}
\end{figure}

\begin{equation}
\begin{split}
\frac{d\sigma}{dx}=\frac{4\alpha^{3}}{3E_{1}m_{e}^{2}x}\left[(x(3x-4)+4)\times\right.\\
\left.\times\left(Z\ln\left(\frac{1194}{Z^{2/3}}\right)+Z^{2}\ln\left(\frac{184}{Z^{1/3}}\right)\right)-\frac{(x-1)\left(Z+Z^{2}\right)}{3}\right],\label{eq:normalBS}
\end{split}
\end{equation}
Note that in the massless case, the term proportional to $Z+Z^{2}$ is often
neglected, whereas in the massive case, it is the dominant contribution
to the cross-section. The $x\rightarrow0$ divergence for massless
photons is obviously removed by the photon mass and the expression
for massive photons is peaked at 1. What is more important, for interesting
values of $m_{\gamma}$, the suppression of the cross-section by the
photon mass is huge for almost every $x$, with a typical value of
$m_{\gamma}$ as shown in Figure \ref{fig:250-500-1}.

\section{Simulations}

As the cross-section for dark photon production is very small, we
can incorporate this effect into full Monte Carlo simulations of the
air showers in a very simple way: for each bremsstrahlung event in
the shower we give either the ratio of eqs. (\ref{eq:darkBS}) and (\ref{eq:normalBS})
using the kinematics of the particular scattering, or zero when a
dark photon could not be emitted. Then we sum these values over the
whole shower, resulting in an approximate mean number of massive photons
that would be produced in the given shower. This scheme is neglecting the influence of dark
photon production on the rest of the shower and exact energy conservation,
but we assume that the production rate is small compared to the overall number of particles in the shower and thus such correction will be very small. 
We will check the validity of this assumption after presenting the results.

Specifically, we use the CORSIKA \cite{Corsika:1998} program (version 6.900) for detailed simulation of extensive air showers. The EGS4 routines are used for the electromagnetic cascade, as they produce the necessary data for individual particles, as opposed to the analytical NKG package. The EGS4 also incorporates the LPM (Landau-Pomeranchuk-Migdal) effect \cite{LPM}. The Earth's magnetic field and altitude were adjusted to the Pierre Auger Observatory site. Hadronic interactions were primarily treated by GHEISHA \cite{geisha} (low-energy) and QGSJET II \cite{Ostapchenko200873} (high-energy) models and simulations were carried out with primary proton and iron particles at primary energies $10^{18}$~eV, $10^{19}$~eV and $10^{20}$~eV respectively, at zenith angle 38$^{\circ}$ as the most common arrival direction for a detector located at a flat surface. The azimuth is generated randomly. To estimate the effect of the choice of hadronic models, we compare the results at one chosen primary particle type (proton) and energy ($10^{19}$~eV) with simulations using either FLUKA \cite{FLUKA} for the low-energy hadronic interactions or EPOS 1.99 \cite{EPOS} or SIBYLL 2.1 \cite {SIBYLL} for the high-energy interactions. Together we carried out 9 sets of simulations with different settings with 100 simulations in each set.

\section{Results}

\begin{table}
\centering
\caption{Number of dark photons produced in a shower using simulations with GEISHA and QGSJET II for proton and iron primary particles and different energies}
\begin{tabular}{l c c c c}
&\multicolumn{2}{c}{\bf{proton}}&\multicolumn{2}{c}{\bf{iron}} \\
\bf{Energy}&mean value&central 68 \%&mean value&central 68 \%\\
\hline
$10^{18}$~eV&$0.39\pm0.02$&0.30--0.45&$0.28\pm0.03$&0.20--0.32\\
$10^{19}$~eV&$1.05\pm0.06$&0.60--1.34&$1.3\pm0.2$&0.38--1.33\\
$10^{20}$~eV&$5.9\pm2.7$&2.5--7.4&$5.8\pm1.2$&1.6--7.6\\
\hline

\end{tabular}

\end{table}

\begin{table}
\centering
\caption{Number of dark photons produced in a shower induced by a proton at $10^{19}$~eV with different low- and high-energy hadronic interaction models}
\begin{tabular}{l c c }
\bf{Interaction models}&mean value&central 68 \%\\
\hline
GEISHA+QGSJETII&$1.05\pm0.06$&0.60--1.34\\
FLUKA+QGSJETII&$1.40\pm0.20$&0.64--1.68\\
GEISHA+SIBYLL&$1.50\pm0.23$&0.76--1.61\\
GEISHA+EPOS&$1.06\pm0.08$&0.59--1.47\\
\hline

\end{tabular}

\end{table}

To give concrete numbers, we choose the mass of the photon to be 250~MeV as a favourable value for the muon production. From Fig.~1 one can see that the production of dark photons is larger for smaller values of photon mass,  but when we want to consider muon production, the obvious lower limit is 212~MeV (twice the muon mass) -- we choose a value slightly higher to avoid dealing with threshold effects; incidentally it is the value considered as likely in  \cite{ArkaniHamed:2008qn} based on astrophysical data.

In Table~1 we present the average numbers of dark photons (per shower) produced in simulations with GEISHA and QGSJETII with different primary energies and particle type.  The differences between individual simulations are a combination of physical fluctuations of the interactions, known as the ``shower-to-shower fluctuations" \cite{StS}, and of the effect of the particle thinning \cite{Hillas}. The distributions of these fluctuations are not Gaussian -- in fact, they follow approximately log-normal distributions and thus the mean values and their uncertainties are heavily influenced by the tails of the distributions. To give the reader a better idea of the fluctuations, we have indicated the range of values in which the central 68 \% (corresponding to 1 $\sigma$ for the normal distribution) of each distribution lies, instead of just the standard deviation. The increase of the fluctuations between simulations (and thus of the uncertainty of the mean value) with energy is related to the effect of thinning which is set relatively to the primary energy and thus it is a stronger effect at higher energy (otherwise the computing times would be prohibitively large). While the distribution of the results from individual simulations is not normal, we assume that the distribution of the means of different samples of a given size approaches the normal distribution and thus we estimate the error of the mean as the standard deviation of each sample divided by square root of the number of simulations. In Table~2 we present for one particular energy and primary particle a comparison between results obtained using different low- and high-energy hadronic interaction models.

All these results have to be further multiplied by the square of the suppression factor $\epsilon$ introduced in Sect.~1 as there is one vertex with dark photon and normal matter in the Feynman diagram for Bremsstrahlung and the matrix element is squared in the cross-section. The exact value of $\epsilon$ is to some extent a free parameter of the model, thus we present the results without
it. Note that to avoid direct detection and cosmological constraints, 
this factor has to be of the order $\epsilon^{2}\approx10^{-4}$ or even
lower.  A dark photon with $m_{\gamma}=250$~MeV decays to a pair of muons almost exactly in 50 \% of cases, so there is on average one muon ($\mu^+$ or $\mu^-$) produced per one dark photon. 

\section{Discussion}

Both the composition of the cosmic rays at ultra-high energies and the correct choice of a hadronic interaction model are currently unknown. Nevertheless we observe that, at our comparison energy of $10^{19}$~eV, all the values obtained for different choices of both composition and hadronic model are compatible with each other within less than 2 standard errors.\footnote{The relatively large difference between SIBYLL and other models may very well be just a fluctuation. Alternatively, it could be due to lower multiplicity of high-energy interactions in SIBYLL (see  \cite {SIBYLL}). The lower multiplicity causes higher average energies of secondary particles which then go through more generations of interactions before decaying. Each such generation feeds the electromagnetic component of the shower via $\pi^0$ decays, thus increasing the number of photons available for bremsstrahlung. } Thus our analysis is largely independent of these unknown inputs. The amount of dark photons produced is dependent on the primary energy, but this result is expected as even with a relatively simple model \cite{MH} it can be shown that the overall number of particles in a shower scales approximately linearly with primary energy. For comparison with our values, note that total number of muons on the ground level for a single shower in our set of simulations is of the order of $10^{7}$ -- $10^{9}$ (depending mainly on primary energy and composition). 

Even at the highest energy observed in cosmic rays ($10^{20}$~eV) and using the maximal value of  $\epsilon^{2}=10^{-4}$, we predict less than one muon originating from a dark photon to be produced in 1000 showers.  As the flux of primary particles of such energy is very small (about 0.01 particle per km$^2$ per year \cite{spektrum}) even the current largest UHECR detector, The Pierre Auger Observatory, has not yet observed that many events at this energy (in 2011 they reported less than a hundred events above $5\times10^{19}$~eV \cite{augerE}). The situation is slightly better at lower energies: for $10^{18}$~eV (again using the maximal value of $\epsilon$) our results permit only one muon from a dark photon in approximately 30000 showers, but as the energy spectrum is very steep, the flux is much larger,  $\sim$100 primary particles per km$^2$ per year. Still, it is obvious that the influence of possible dark photon production on the muon content in extensive air showers is extremely small.

From the previous discussion it also follows that the assumption in our method (that the effect of the dark photon production on the simulation of the rest of the shower is small) is justified, as in the vast majority of showers, there are no dark photons produced at all.

\section{Conclusions}

Motivated by recent observational and theoretical development in describing the possible dark matter in our Universe on one side and the discrepancy between the observed muon number in extensive air showers with simulations on the other, we have conducted a study where we tried to see if a particular feature of the former (the production of dark photons) could help the fix the discrepancy in the latter. We found out that even for the most favourable values of parameters that can be feasibly adopted ($m_{\gamma}=250$~MeV and $\epsilon^{2}=10^{-4}$) and for various energies, primary particles and interaction models, the muon production in the EAS caused by dark photon decay is negligible. This result is valid for any massive photon-like particle with an interaction that is governed by the standard quantum electrodynamics, modified only by the suppression factor $\epsilon$. As an useful by-product, we presented a closed-form expression for the bremsstrahlung of a massive photon from a lepton in the framework of the Weizs\"{a}cker-Williams approximation.

\section*{Acknowledgements}

We are grateful to Jan \v{R}\'{i}dk\'{y} for the idea to consider these models in extensive air showers and for his valuable advice. The contribution is prepared with the support of Ministry of Education, Youth and Sports of the Czech Republic within the project LA08016 and with the support of the Charles University in Prague within the project 119810.

\section*{References}

\bibliographystyle{elsarticle-num}
\bibliography{darphoton}

\end{document}